\documentstyle[aps
,preprint%
]{revtex}
\newcommand{\be}{\begin{equation}}
\newcommand{\ee}{\end{equation}}
\newcommand{\bea}{\begin{eqnarray}}
\newcommand{\eea}{\end{eqnarray}}
\newcommand{\nn}{\nonumber \\}

\newcommand{\Tr}{{\rm Tr}}
\begin{document}
\draft
\preprint{AJC-HEP-27}
\date{\today
}
\title{
Statistical Mechanics of Charged Particles
in Einstein-Maxwell-Scalar Theory
}
\author{Kiyoshi~Shiraishi%
\thanks{e-mail: {\tt g00345@sinet.ad.jp, shiraish@air.akita-u.ac.jp}
}}
\address{Akita Junior College\\
Shimokitade-sakura, Akita-shi, Akita 010, Japan
}
\author{Takuya~Maki%
\thanks{e-mail: {\tt maki@nls.kitasato-u.ac.jp}
}}
\address{Kitasato University\\
Kitasato, Sagamihara-shi, Kanagawa 228, Japan
}
\maketitle
\begin{abstract}
We consider an $N$-body system of charged particle coupled to
gravitational, electromagnetic, and scalar fields.
The metric on moduli space for the system can be considered if
a relation among the charges and mass is satisfied, which
includes the BPS relation for monopoles and the extreme
condition for charged dilatonic black holes.
Using the metric on moduli space in the long distance approximation,
we study the statistical mechanics of the charged particles
at low velocities.
The partition function is evaluated as the leading order
of the large $d$ expansion, where $d$ is the spatial dimension of the
system and will be substituted finally as $d=3$.
\end{abstract}
\vfill
\eject


Solitons in field theories have attracted
much attention in past two decades.
The dynamics of solitons have been studied
not only by numerical methods
but also by analytical methods.
If a multi-soliton solution saturates
a certain Bogomol'nyi-type bound,
there is no net static force among the solitonic objects described
by such a solution.
The existence of various types of forces is essential
for the balanced situation.
The low-energy interaction of such solitons can be described
in terms of the space of parameters
in static solutions (moduli space)~\cite{Manton}.

Recently, the statistical mechanics of vortices in $(2+1)$ dimensions
was studied by Manton and Shah~\cite{MS,Shah} under the assumption
that the dynamics of vortices is well approximated by the geodesic
motion on moduli space.
In the present paper, we investigate thermodynamic behavior of the
gas of the charged particles interacting by
gravitational, electromagnetic, and scalar forces.
This model can be regarded as a point particle limit of
monopoles and/or black holes
when there is a special relation among the charges and mass.
We adopt the lagrangian of
particles at low energies and at large separations
to compute the partition function of the system.


Suppose the action for the $N$ particles of mass~$m_{\alpha}$,
electric charge~$q_{\alpha}$
and scalar charge~$\sigma_{\alpha}$ in the $(d+1)$
dimensional spacetime $(d\geq 3)$ is
given by
\bea
L=&-&\sum_{\alpha=1}^{N}\int ds_{\alpha}
\left(m_{\alpha}-\sigma_{\alpha}\varphi-
q_{\alpha}A_{\mu}\frac{\partial{x_{\alpha}^{\mu}}}{\partial{s_{\alpha}}}
\right)\nn
&+&\int d^{d}x\ \frac{\sqrt{-g}}{16\pi}
\left[\frac{1}{G}R-2(\nabla\varphi)^{2}-F^{2}
\right]+
\mbox{(surface terms)}\ ,
\eea
where $R$ is the scalar curvature and $\varphi$ is a real scalar field.
The electromagnetic field $A_{\mu}$ is related to
 the field strength $F_{\mu\nu}$ by
$F_{\mu\nu}=\partial_{\mu}A_{\nu}-\partial_{\nu}A_{\mu}$,
where $\mu$, $\nu$ run over $0,1,\ldots,d+1$.
$G$ is the Newton's constant. We will finally focus on the case of
$d=3$, but for later use in an approximation scheme, we leave $d$ as
an arbitrary integer for a while.

One of the present authors has obtained the low-energy lagrangian
including Li\'enard-Wiechert potential in the large mutual
distance limit~\cite{Shi1,Shi2}.
Using abbreviated notations
\bea
e_{\alpha\beta}^{2}\equiv
q_{\alpha}q_{\beta}-\sigma_{\alpha}\sigma_{\beta}-
\frac{2(d-2)}{d-1}Gm_{\alpha}m_{\beta}\ ,\\
\gamma^{2}_{\alpha\beta}\equiv
\left[\sigma_{\alpha}\sigma_{\beta}-\frac{2d}{d-1}Gm_{\alpha}m_{\beta}
\right]\ ,
\eea
we can write the effective lagrangian for the particles as
\bea
L&=&\frac{1}{2}\sum_{\alpha=1}^{N}m_{\alpha}\vec{v}_{\alpha}^{2}\nn
&-&\sum_{\alpha<\beta}\frac{4\pi\gamma_{\alpha\beta}^{2}}{A_{d-1}}
\frac{|\vec{v}_{\alpha}-\vec{v}_{\beta}|^{2}}
{2(d-2)\ r_{\alpha\beta}^{d-2}}\nn
&+&\sum_{\alpha<\beta}\frac{4\pi e_{\alpha\beta}^{2}}{A_{d-1}}\frac{
\left[\vec{v}_{\alpha}\cdot\vec{v}_{\beta}+
(d-2)(\vec{n}_{\alpha\beta}\cdot\vec{v}_{\alpha})
(\vec{n}_{\alpha\beta}\cdot\vec{v}_{\beta})
\right]}{2(d-2)\ r_{\alpha\beta}^{d-2}}\nn
&-&\sum_{\alpha<\beta}\frac{4\pi e_{\alpha\beta}^{2}}{A_{d-1}}
\frac{1}{(d-2)\ r_{\alpha\beta}^{d-2}}\nn
&+&\cdots\ ,
\label{eq:La}
\eea
where $A_{d-1}=\frac{2\pi^{d/2}}{\Gamma (d/2)}$.%
\footnote{We use a different unit system from that
in ref.~\cite{Shi1,Shi2} and fixed coefficients.}
 $r_{\alpha\beta}$ is the distance between two point sources
denoted by $\alpha$ and $\beta$, and $\vec{n}_{\alpha\beta}$
stands for the unit normal vector
in the direction $\beta$-$\alpha$.
Here, the triple dots ($\cdots$) contain $O(v^{3})$
 and $O(1/r^{2(d-2)})$ interactions.
The radiation reaction is negligible at low velocities.
Note that this lagrangian can be utilized for the system of magnetic
monopoles,
where the duality relation holds.

In eq.~(\ref{eq:La}), the static interaction is cancelled out if
\be
e_{\alpha\beta}^{2}=0\ .
\label{eq:balance}
\ee
For the BPS monopoles in $(3+1)$ dimensions,
if gravity is negligible ($G\rightarrow 0$),
the relation $q_{\alpha}=\sigma_{\alpha}$ holds for
each monopoles~\cite{GM};
 then the balance
condition~(\ref{eq:balance}) is satisfied.
For extreme dilatonic black holes~\cite{GHS,Shi1,Shi2},
\bea
q_{\alpha}&=&\sqrt{\frac{2(d-2+a^{2})}{d-1}}\sqrt{G}m_{\alpha}\ ,\nn
\sigma_{\alpha}&=&a\sqrt{\frac{2}{d-1}}\sqrt{G}m_{\alpha}\ ,
\eea
where $a$ is the dilaton coupling constant.
The balance condition~(\ref{eq:balance}) is satisfied also in this case.

When the balance condition~(\ref{eq:balance}) is satisfied,
there remains the interaction term which depends on the relative
velocities.
In this critical case, the slow motion of the well-separated particles
is geodesic on the moduli space, and
the evaluation of the partition function for the system
of the particles reduces to the volume of the moduli space~\cite{MS}.
For non-critical cases, the deviation from the geodesic motion is
expected to be modest if the couplings of the static potential
and the other complicated velocity-dependent interaction are
small~\cite{Shah}.
Therefore we assume that we can express the partition function of the
system in terms of a collective coordinates on the moduli space
in the case of general charges and mass
at least in the leading order calculation.


The lagrangian of the $N$-body system can be read as
\be
L=\frac{1}{2}g_{AB}\dot{Q}^{A}\dot{Q}^{B}-{\sf V}(Q)\ ,
\ee
where $\{A\}=\{\alpha,i\}$ with
$\alpha=1,2,\ldots,N$ and $i=1,2,\ldots,d$.
$Q$ is considered as the coordinates in $Nd$ dimensional space.
$g_{AB}$ is called as the metric on moduli space of $N$-body system
when the balance condition is satisfied.
{\sf V}(Q) denotes the static potential.

The partition function is then given by
\be
Z=\frac{1}{h^{Nd}}\int [dQdP]\ e^{-\beta H(P,Q)}\ ,
\ee
where
$h=2\pi\hbar$, $\beta=(k_{B}T)^{-1}$; $T$ is the temperature of the
system and $k_{B}$ is the Boltzmann's constant.

The momentum corresponding to $Q$ is defined as
\be
P_{A}=\frac{\partial{L}}{\partial{\dot{Q}^{A}}}=g_{AB}\dot{Q}^{B}\ .
\ee
Using this canonical momentum, the Hamiltonian is expressed as
\be
H=\frac{1}{2}g^{AB}P_{A}P_{B}+{\sf V}(Q)\ ,
\ee
where $g^{AB}$ satisfies $g^{AC}g_{CB}=\delta_{B}^{A}$.

Performing the Gaussian integrals over $P$, we find
\be
Z=\left(\frac{2\pi}{\beta h^2}\right)^{\frac{Nd}{2}}\
\int [dQ]\ \sqrt{\det\ g_{AB}(Q)}
\exp\left[-\beta\ {\sf V}(Q)\right]\ .
\label{eq:Z}
\ee
If the balance condition is satisfied, the partition function $Z$ can be
expressed by using the volume of the moduli space ${\cal V}_{\cal M}$ as
\be
Z=\left(\frac{2\pi}{\beta h^2}\right)^{\frac{Nd}{2}}\ {\cal V}_{\cal M}\ ,
\ee
where
\be
{\cal V}_{\cal M}=\int [dQ]\ \sqrt{\det\ g_{AB}}\ .
\ee


We turn now to the evaluation of the partition function
for the particle system governed
by gravitational, electromagnetic, and scalar forces.
Keeping terms of order $v^2$ and $1/r^{d-2}$,
we recast the classical lagrangian for $N$ identical particles of equal
charges~$q$,~$\sigma$ and mass~$m$ as
\bea
L&=&\frac{1}{2}m\sum_{\alpha=1}^{N}\vec{v}_{\alpha}^{2}\nn
&-&\frac{4\pi\gamma^{2}}{A_{d-1}}\sum_{\alpha<\beta}
\frac{|\vec{v}_{\alpha}-\vec{v}_{\beta}|^{2}}
{2(d-2)\ r_{\alpha\beta}^{d-2}}\nn
&+&\sum_{\alpha<\beta}\frac{4\pi e^{2}}{A_{d-1}}\frac{
\left[\vec{v}_{\alpha}\cdot\vec{v}_{\beta}+
(d-2)(\vec{n}\cdot\vec{v}_{\alpha})(\vec{n}\cdot\vec{v}_{\beta})
\right]}{2(d-2)\ r_{\alpha\beta}^{d-2}}\nn
&-&\sum_{\alpha<\beta}\frac{4\pi e^{2}}{A_{d-1}}
\frac{1}{(d-2)\ r_{\alpha\beta}^{d-2}}\ ,
\label{eq:L}
\eea
where $\gamma^{2}\equiv
\sigma^{2}-\frac{2d}{d-1}Gm^{2}$
and
$e^{2}\equiv
q^{2}-\sigma^{2}-\frac{2(d-2)}{d-1}Gm^{2}$.

We use this lagrangian for well-separated
particles to calculate the partition function.
The approximation is justified if the average distance
between particles is large. Thus we have to consider a dilute gas
of the particles in our analysis.

{}From the lagrangian (\ref{eq:L}), we obtain the metric
$g_{AB}$ as follows:
\be
g_{AB}=m\ I_{AB}-\gamma^{2}\ U_{AB}+e^{2}\ V_{AB}\ ,
\ee
with
\bea
I_{AB}&=&\delta_{ij}\otimes\delta_{\alpha\beta}\ ,\\
U_{AB}&=&\delta_{ij}\otimes u_{\alpha\beta}\ ,
\eea
where
\be
u=\left(
\begin{array}{ccccc}
\sum_{\alpha\neq 1}^{N}\phi_{1\alpha} & -\phi_{12} & -\phi_{13} & \cdots
& -\phi_{1N} \\
-\phi_{12} & \sum_{\alpha\neq 2}^{N}\phi_{2\alpha} & -\phi_{23} & \cdots
& -\phi_{2N} \\
-\phi_{13} & -\phi_{23} & \sum_{\alpha\neq 3}^{N}\phi_{3\alpha} & \cdots
& -\phi_{3N} \\
\vdots     & \vdots     &  \vdots     & \ddots   & \vdots \\
-\phi_{1N} & -\phi_{2N} & -\phi_{3N} & \cdots &  \sum_{\alpha\neq N}^{N}
\phi_{N\alpha}
\end{array}
\right)\ ,
\ee
\be
V_{AB}=
\left\{
\begin{array}{cl}
\left(\delta_{ij}+n_{\alpha\beta,i}n_{\alpha\beta,j}\right)\
\phi_{\alpha\beta}&(\alpha\neq\beta)\\
0&(\alpha=\beta)
\end{array}
\right.\ ,
\ee
and
\be
\phi_{\alpha\beta}=\frac{4\pi}{A_{d-1}(d-2)\ r_{\alpha\beta}^{d-2}}\ ,
\ee

First we consider the case that $e^{2}=0$.
Then we get
\bea
\sqrt{\det g_{AB}}&=&m^{\frac{Nd}{2}}\exp\frac{1}{2}\Tr\ln
(I_{AB}-\frac{\gamma^{2}}{m}U_{AB})\nn
&=&m^{\frac{Nd}{2}}\exp\frac{d}{2}\Tr\ln
(\delta_{\alpha\beta}-\frac{\gamma^{2}}{m}u_{\alpha\beta})\nn
&=&m^{\frac{Nd}{2}}\exp\left\{-\frac{d}{2}
\left[\frac{\gamma^{2}}{m}\Tr\ u+\frac{1}{2}
\left(\frac{\gamma^{2}}{m}\right)^{2}\Tr\ u^{2}+\cdots\right]\right\}\ .
\label{eq:det}
\eea

Taking the limit
$d\rightarrow \infty$,
$\frac{d}{m}\equiv\frac{1}{\tilde{m}}$ fixed,%
\footnote{At the same time, we have to set
$Gm^{2}=\tilde{G}\tilde{m}^{2}$.}
one finds that $\frac{1}{\tilde{m}^{2}}\phi^{2}$ contribution
from $\Tr\ u^{2}$ is
suppressed by a factor $1/d$ in comparison with that from $\Tr\ u$
in the expression~(\ref{eq:det}), and also higher order terms
of $\frac{1}{\tilde{m}}\phi$ from $\Tr\ u^{n}\ \ (n\geq 2)$
or the product of them are more or so suppressed.
Thus, according to this ``large $d$'' approximation,
the expression~(\ref{eq:det}) reduces to
\bea
\sqrt{\det g_{AB}}&\approx&m^{\frac{Nd}{2}}\exp
\left[-\frac{\gamma^{2}d}{2m}\Tr\ u\right]\nn
&=&m^{\frac{Nd}{2}}\exp\left(-\frac{\gamma^{2}d}{m}
\sum_{\alpha<\beta}\phi_{\alpha\beta}\right)\nn
&=&m^{\frac{Nd}{2}}\exp\left(-\frac{\gamma^{2}d}{m}
\sum_{a=1}^{M}\phi_{a}\right)\ ,
\eea
where $a$ represents the pair of indices $\{\alpha,\beta\}$,
 such as $\{1,2\}$,
and $M=N(N-1)/2$.

In this leading $1/d$ approximation scheme,
the contribution of $V_{AB}$ is neglected even for finite $e^{2}$,
since $V_{AB}$ is traceless.


One can find that
this closely resemble the contribution of static potential
to the partition function,%
\footnote{Thus positive $\gamma^{2}$ give rise to an effective
 repulsive force between the particles.}
 which can be written by
\be
\exp\ (-\beta\ {\sf V})=
\exp\left(-e^{2}\beta
\sum_{a=1}^{M}\phi_{a}\right)\ .
\ee


Thus the integral over $Q$ in eq.~(\ref{eq:Z}) can be written
in a combined form
\be
\frac{V^{N}}{N!}m^{\frac{Nd}{2}}e^{NW}\ ,
\ee
where $V$ is the volume of the system.
As the thermodynamic limit $N\rightarrow\infty$ and
$V\rightarrow\infty$ is taken, $W$ is given by~\cite{Abe}
\be
NW=\frac{1}{V^{N}}
{\sum_{n_{1},\ldots,n_{M}}}^{'}
\left(-\frac{\gamma^{2}d}{m}- e^{2}\beta\right)^{\sum_{a}^{M}n_{a}}
\frac{\int (\phi_{1}^{n_{1}}\phi_{2}^{n_{2}}\cdots
\phi_{M}^{n_{M}})_{irr}\prod_{\alpha=1}^{N}d^{d}x_{\alpha}}%
{n_{1}! n_{2}! \cdots n_{M}!}\ ,
\ee
where $irr$ stands for the irreducible set and
$\sum^{'}$ means the sum
excluding the term of $n_{1}=n_{2}=\cdots=n_{M}=0$.

Here we will use the ring approximation~\cite{Abe};
we pick up the irreducible set of a ring shape.
Using this approximation with the Fourier transform, we get
\bea
W_{0}&=&\frac{1}{2V}\sum_{n=2}^{\infty}
\frac{(-\gamma^{2}d-e^{2}\beta m)^{n}\rho^{n-1}}{m^{n}n}
\sum_{\vec{k}}\left(\frac{4\pi}{|\vec{k}|^{2}}\right)^{n}\nn
&=&\frac{1}{2\rho V}\sum_{\vec{k}}\left\{-\ln
\left[1+\rho\frac{\kappa^{2}}{|\vec{k}|^{2}}\right]+
\rho\frac{\kappa^{2}}{|\vec{k}|^{2}}\right\}\ ,
\eea
where the density
of particles is defined as $\rho=N/V$ and
$\kappa^{2}=4\pi\left(\frac{\gamma^{2}d}{m}+ e^{2}\beta\right)\rho$.
Here the suffix $0$ means that we used the ring approximation.


Consequently, we have the result for $d=3$, though $d$ may not be
very big:
\bea
Z&\approx&\frac{V^{N}}{N!}
\left(\frac{2\pi m}{\beta h^2}\right)^{\frac{3N}{2}}\ e^{NW_{0}}\nn
&=&\frac{V^{N}}{N!}
\left(\frac{2\pi m}{\beta h^2}\right)^{\frac{3N}{2}}
\exp\left\{\frac{N}{12\pi}\left[\frac{4\pi}{m}(3\gamma^{2}+e^{2}\beta m)
\right]^{\frac{3}{2}}
\rho^{\frac{1}{2}}\right\}\ .
\eea

This yields the equation of state given by
\be
\frac{pV}{Nk_{B}T}=1-
\frac{1}{24\pi}\left[\frac{4\pi}{m}(3\gamma^{2}+e^{2}\beta m)
\right]^{\frac{3}{2}}
\rho^{\frac{1}{2}}\ ,
\label{eq:DH}
\ee
where $p$ is the pressure.


The deviation from an ideal gas is independent of the temperature
for $e^{2}=0$.
This is true if one may use any approximation, because the partition
function is proportional to the volume of the moduli space
${\cal V_{M}}$ in this special case.
One can also see from eq.~(\ref{eq:DH}) that
a possible phase transition occurs at high density,
but the large distance approximation
may become unreliable in such a case.

More interesting aspect is
the appearance of the instability of the system.
In the case of $\kappa^{2}<0$,
imaginary contributions appear in the free energy ($\propto\ln{Z}$),
which indicates the instability or the absence of
the thermodynamical equilibrium in the system.
The ``phase'' diagram for
charges is shown in FIG.~\ref{fig1},
where the possible sign of $\kappa^{2}$ is exhibited.
When the charges fall into the upper-right region,
the system is stable for any temperature.
In the lower-left region,  the system is always unstable.
In the upper-left region,  the system becomes
unstable at low temperature.
In the lower-right region, the system becomes
unstable at high temperature.

For charged dilaton black holes~\cite{GHS} with $a^{2}<3$,
the allowed charges fall into the region
that $\kappa^{2}<0$ for any value of $\beta m$ in the diagram.
The instability may be considered as a characteristic feature
of gravitating systems.
For $a^{2}>3$, if $\gamma^{2}>0$ is satisfied, $\kappa^{2}$ becomes
positive when the temperature is sufficiently high.%
\footnote{Actually, the charges of the any dilatonic black holes
fall into the region between the lines $q^{2}=\sigma^{2}+Gm^{2}$
and $q^{2}=\sigma^{2}$.}

The charges of the {\em extreme} dilaton black holes correspond to
the line described by $q^{2}=\sigma^{2}+Gm^{2}$.
The system of the extreme black holes is always unstable for $a^{2}<3$,
while for $a^{2}>3$ the system is stable according to the present
analysis.
The extreme case for $a^{2}=3$ corresponds to
the point $K$ in FIG.~\ref{fig1}.
This is also corresponding to the Kaluza-Klein monopole~\cite{Ruback}.
In this case, since there is no interaction of the order of $v^2$,
the system is equivalent to the classical ideal gas of free particles.
On the other hand, the extreme Reissner-Nordstr\"om black hole
corresponds
to the point $E$~\cite{GR}. In this case,
the system is essentially unstable.

One must also bear in mind that the present estimation of
the partition function has been obtained by using several approximations.
More precise approximation schemes should be studied for some
specific models of extended objects as well as of the point particles.
In the case that the minute structure of moduli space is well-known,
such as for the BPS monopoles~\cite{Manton,BPS}
and charged black holes~\cite{FerEar,Shi2,Shi3},
application of the moduli metric to
the thermodynamics requires further work
with and without quantum mechanical considerations.

Finally, we expect that the result of the work by Gibbons and
Manton~\cite{GM}
on the moduli space of well-separated dyons may be utilized
to study the thermodynamics of dyons. Further investigation will be
reported
elsewhere.



\newpage
\begin{center}
\begin{figure}
\unitlength 1mm
\begin{picture}(80,80)(0,-10)
\thicklines
\put(0,0){\vector(0,1){60}}
\put(0,0){\vector(1,0){80}}
\put(10,0){\line(0,1){1}}
\put(20,0){\line(0,1){1}}
\put(30,0){\line(0,1){1}}
\put(40,0){\line(0,1){1}}
\put(50,0){\line(0,1){1}}
\put(60,0){\line(0,1){1}}
\put(70,0){\line(0,1){1}}
\put(0,10){\line(1,0){1}}
\put(0,20){\line(1,0){1}}
\put(0,30){\line(1,0){1}}
\put(0,40){\line(1,0){1}}
\put(0,50){\line(1,0){1}}
\put(0,0){\makebox(0,0)[rt]{$0$}}
\put(80,0){\makebox(0,0)[lt]{$\frac{q^{2}}{Gm^{2}}$}}
\put(0,60){\makebox(0,0)[rb]{$\frac{\sigma^{2}}{Gm^{2}}$}}
\thinlines
\put(10,0){\line(1,1){60}}
\put(0,30){\line(1,0){80}}
\put(60,40){\makebox(0,0)[lb]{\shortstack{$\kappa^{2}>0$\\
for any $\beta m$}}}
\put(5,25){\makebox(0,0)[lt]{\shortstack{$\kappa^{2}<0$\\
for any $\beta m$}}}
\put(5,40){\makebox(0,0)[lb]
{\shortstack{$\kappa^{2}$: positive$\rightarrow$negative\\
as $\beta m$: $0\rightarrow\infty$}}}
\put(40,25){\makebox(0,0)[lt]
{\shortstack{$\kappa^{2}$: positive$\rightarrow$negative\\
as $\beta m$: $\infty\rightarrow 0$}}}
\put(10,-1){\makebox(0,0)[ct]{$1$}}
\put(40,-1){\makebox(0,0)[ct]{$4$}}
\put(-1,30){\makebox(0,0)[rc]{$3$}}
\put(40,30){\circle*{1}}
\put(10,0){\circle*{1}}
\put(40,31){\makebox(0,0)[cb]{$K$}}
\put(10,1){\makebox(0,0)[cb]{$E$}}
\end{picture}
\caption{The phase diagram for charges of the particles, for $d=3$.}
\label{fig1}
\end{figure}
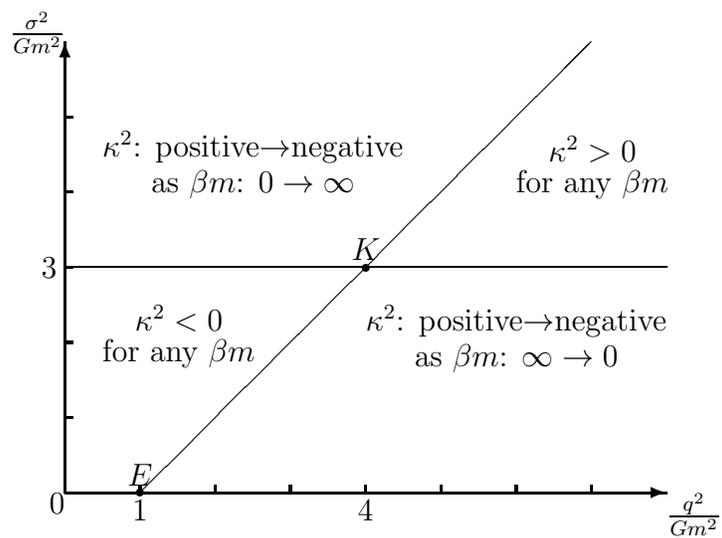
\end{center}

\end{document}